\documentclass[aps,prl,twocolumn,groupedaddress]{revtex4}

\usepackage{graphicx}

\begin{document}

\title{Entanglement in Mesoscopic Structures: 
Role of Projection}

\author{A.V.\ Lebedev$^{\, a}$, G.\ Blatter$^{\, b}$,
C.W.J.\ Beenakker$^{\, c}$, and G.B.\ Lesovik$^{\, a}$}
\affiliation{$^{a}$L.D.\ Landau Institute for Theoretical Physics RAS,
117940 Moscow, Russia}
\affiliation{$^{b}$Theoretische Physik, ETH-H\"onggerberg, CH-8093
Z\"urich, Switzerland}
\affiliation{$^{c}$Instituut-Lorentz, Universiteit Leiden,
P.O.\ Box 9506, 2300 RA Leiden, The Netherlands}

\date{\today}

\begin{abstract}
   We present a theoretical analysis of the appearance of
   entanglement in non-interacting mesoscopic structures.
   Our setup involves two oppositely polarized sources 
   injecting electrons of opposite spin into the two 
   incoming leads. The mixing of these polarized streams 
   in an ideal four-channel beam splitter produces two 
   outgoing streams with particular tunable correlations. 
   A Bell inequality test involving cross-correlated 
   spin-currents in opposite leads signals the presence 
   of spin-entanglement between particles propagating 
   in different leads. We identify the role of fermionic 
   statistics and projective measurement in the 
   generation of these spin-entangled electrons.
\end{abstract}

\maketitle

Quantum entangled charged quasi-particles are perceived as a
valuable resource for a future solid state based quantum
information technology. Recently, specific designs for mesoscopic
structures have been proposed which generate spatially separated
streams of entangled particles 
\cite{ent_sc,ent_qd,chtchelkatchev_02,samuelsson_03}.
In addition, Bell inequality type measurements have been conceived
which test for the presence of these non-classical and non-local
correlations \cite{chtchelkatchev_02,samuelsson_03}. Usually,
entangled electron-pairs are generated through specific
interactions (e.g., through the attractive interaction in a
superconductor or the repulsive interaction in a quantum dot) and
particular measures are taken to separate the constituents in
space (e.g., involving beam splitters and appropriate filters).
However, recently it has been predicted that non-local
entanglement as signalled through a violation of Bell inequality
tests can be observed in non-interacting systems as well
\cite{beenakker_03,fazio_03,samuelsson_04,lebedev_03}. 
The important task then is to identify the origin of 
the entanglement; candidates are the fermionic statistics, 
the beam splitter, or the projection in the Bell 
measurement itself \cite{bosehome_02,samuelsson_04}. 

Here, we report on our study of entanglement in a 
non-interacting system, where we make sure, that 
the particles encounter the Bell setup in a 
non-entangled state. Nevertheless, we find 
the Bell-inequality to be violated and conclude that 
the concomittant entanglement is produced in a wave 
function projection during the Bell measurement.
We note that wave function projection as a resource 
of non-local entanglement is known for single-particle 
sources (Fock states) \cite{bosehome_02}, a scheme working 
for both bosons and fermions. What is different in Refs.\
\onlinecite{beenakker_03,fazio_03,samuelsson_04,lebedev_03} 
and in the present work is that the sources are many-particle 
states in local thermal equilibrium. It is then essential 
that one deals with fermions; wave function projection 
cannot create entanglement out of a thermal source of bosons.
\begin{figure} [h]
   \includegraphics[scale=0.55]{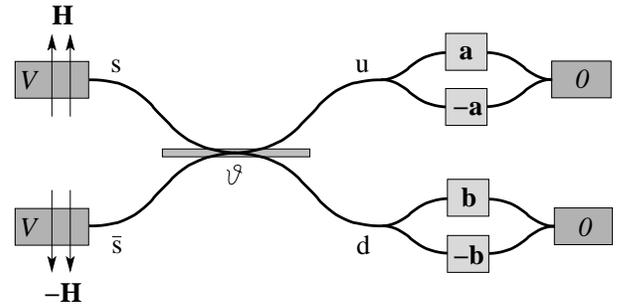}
   \caption[]{Mesoscopic normal-metal structure with a beam splitter
   generating two streams of electrons with tunable correlations
   in the two outgoing arms `u' and `d'. The source (left)
   injects polarized (along the $z$-axis) electrons into the 
   source leads `s' and `$\bar\mathrm{s}$'. The beam splitter
   mixes the two incoming streams with a mixing angle $\vartheta$.
   The scattered (or outgoing) beams are analyzed in a Bell type
   coincidence measurement involving spin-currents projected
   onto the directions $\pm {\bf a}$ (in the `u' lead)
   and $\pm {\bf b}$ (in the `d' lead). The injection reservoirs
   are voltage ($V$) biased against the outgoing reservoirs.
   The Bell inequality test signals the presence of entanglement
   within the interval $|\vartheta - 45^\circ| < 12.235^\circ$.
   We relate this entanglement to the presence of spin-triplet 
   correlations in the projected part of the scattered wavefunction
   describing electron-pairs distributed between the arms.}
   \label{fig:setup}
\end{figure}

The generic setup for the production of spatially 
separated entangled degrees of freedom usually 
involves a source injecting the particles carrying 
the internal degree of freedom (the spin
\cite{ent_sc,ent_qd,fazio_03,lebedev_03} 
or an orbital quantum number 
\cite{samuelsson_03,beenakker_03,samuelsson_04}) and 
a beam splitter separating these particles in space, 
see Fig.\ \ref{fig:setup}. In addition, `filters' may 
be used to inhibit the propagation of unwanted
components into the spatially separated leads
\cite{ent_sc,ent_qd,chtchelkatchev_02,samuelsson_03}, 
thus enforcing a pure flow of entangled particles 
in the outgoing leads. The successful generation of 
entanglement then is measured in a Bell inequality
type setup \cite{BItest}. A surprising new feature 
has been recently predicted with a Bell inequality 
test exhibiting violation in a non-interacting system
\cite{beenakker_03,fazio_03,samuelsson_04,lebedev_03}; 
the question arises as to what produces the entanglement 
manifested in the Bell inequality violation and it is 
this question which we wish to address in the present 
work. In order to do so, we describe theoretically an
experiment where we make sure, that the particles 
are not entangled up to the point where the correlations 
are measured in the Bell inequality setup; nevertheless, 
we find them violated. We trace this violation back to
an entanglement which has its origin in the confluence
of various elements: {\it i)} the Fermi statistics 
provides a noiseless stream of incoming electrons,
{\it ii)} the beam splitter mixes the 
indistinguishable particles at one point in space
removing the information about their origin, {\it iii)}
the splitter directs the mixed product state into the 
two leads thus organizing their spatial separation, 
{\it iv)} a coincidence measurement projects the 
mixed product state onto its (spin-)entangled component 
describing the electron pair split between the two 
leads, {\it v)} measuring the spin-entangled state 
in a Bell inequality test exhibits violation (the 
steps {\it iv)} and {\it v)} are united in our setup).
Note, that the simple fermionic
reservoir defining the source in Ref.\ \onlinecite{lebedev_03}
injects spin-entangled pairs from the beginning; hence an analysis
of this system cannot provide a definitive answer on the minimal
setup providing spatially separated entangled pairs since both the
source and/or the projective Bell measurement could be
responsible for the violation.

Below, we pursue the following strategy: We first define a
particle source and investigate its characteristic via an analysis
of the associated two-particle density matrix. We then define the
corresponding pair wave function (thus reducing the many body
problem to a two-particle problem) and determine its concurrence
following the definition of Schliemann {\it et al.}
\cite{schliemann_01} for indistinguishable particles
(more generally, one could calculate the Slater rank of
the wave function, cf.\ Ref.\ \onlinecite{schliemann_01};
here, we deal with a four-dimensional one-particle Hilbert
space where the concurrence provides a simple and quantitative
measure for the degree of entanglement). For our specially
designed source we find a zero concurrence and hence our
incoming beam is not entangled. We then go over to the
scattering state behind the (tunable) beam splitter and
reanalyze the state with the help of the two-particle
density matrix. We determine the associated two-particle wave
function and find its concurrence; comparing the results for
the incoming and scattered wave function, we will see that the
concurrence is unchanged, a simple consequence of the unitary
action of the beam splitter. However, the mixer removes
the information on the origin of the particles, thus 
preparing an entangled wave function component in the 
output channel. Third, we analyze the component of 
the wave function to which the Bell setup is
sensitive and determine its degree of entanglement;
depending on the mixing angle of the beam splitter, 
we find concurrencies between 0 (no entanglement) and unity
(maximal entanglement). Finally, we determine the violation 
of the Bell inequality as measured through time-resolved 
spin-current cross-correlators and find agreement between 
the degree of violation and the degree of entanglement 
of the projected state as expressed through the concurrence.

Our source draws particles from two spin-polarized reservoirs
with opposite polarization directed along the $z$-axis.
The polarized electrons are injected into source leads `s'
and `$\bar{\rm s}$' and are subsequently mixed in a tunable
four-channel beam splitter, see Fig.\ \ref{fig:setup}.
The outgoing channels are denoted by `u' (for the upper lead)
and `d' (the `down' lead). The spin-correlations in the
scattering channels `u' and `d' are then analyzed in a Bell
inequality test. The polarized reservoirs are voltage biased with
$e V =\mu_{\rm\scriptscriptstyle B} H/2$ equal to the magnetic
energy in the polarizing field $H$; the incoming electron streams
then are fully polarized (the magnetic field is confined to the
reservoirs).

The spin-correlations between electrons in leads `x' and `y'
are conveniently analyzed with the help of the two-particle
density matrix (or pair correlation function)
\begin{equation}
   g_{\vec \sigma}^\mathrm{xy}(x,y)=\mathrm{Tr}
   \bigl(
   \hat{\rho}\hat \Psi_{\mathrm{x}\sigma_1}^\dagger(x)
   \hat \Psi_{\mathrm{y}\sigma_2}^\dagger(y)
   \hat \Psi_{\mathrm{y}\sigma_3}(y)\hat \Psi_{\mathrm{x}\sigma_4}(x)
   \bigr)
   \label{DM}
\end{equation}
with trace over states of the Fermi sea.
Here, $\hat \Psi_{\mathrm{x}\sigma}$ are field operators 
describing electrons with spin $\sigma$ in lead `x' 
and $\hat{\rho}$ is the density operator. The pair 
correlation function (\ref{DM}) is conveniently 
expressed through the one-particle correlators 
$G_{\sigma\bar\sigma}^\mathrm{xy}(x,y) \equiv
\langle\hat \Psi_{\mathrm{x}\sigma}^\dagger(x) \hat
\Psi_{\mathrm{y}\bar\sigma}(y)\rangle$,
\begin{equation}
   g_{\vec \sigma}^\mathrm{xy}(x,y)\!
   =\!G_{\sigma_1\!\sigma_4}^\mathrm{xx}\!(0)
   G_{\sigma_2\!\sigma_3}^\mathrm{yy}\!(0)-
   G_{\sigma_1\!\sigma_3}^\mathrm{xy}\!(x\!-\!y)
   G_{\sigma_2\!\sigma_4}^\mathrm{yx}\!(y\!-\!x).
   \label{DM_G}
\end{equation}
The one-particle correlators can be written in terms of
a product of orbital- and spin parts, $G_{\sigma
\bar\sigma}^\mathrm{xy}(x,y) =G^\mathrm{xy}(x,y)\chi^\mathrm{xy}
(\sigma,\bar\sigma)$, and split into equilibrium and excess terms,
\begin{equation}
   G_{\sigma\bar\sigma}^\mathrm{xy}(x,y)=G_\mathrm{eq}(x,y)
   \chi_\mathrm{eq}^\mathrm{xy}(\sigma,\bar\sigma)+
   G_\mathrm{ex}(x,y)\chi_\mathrm{ex}^\mathrm{xy}(\sigma,\bar\sigma),
   \label{one_p}
\end{equation}
with $G_\mathrm{ex}(x,y)$ vanishing at zero voltage $V$ and zero
polarization field $H$.

In order to find the two-particle density matrix in the
source leads `s', `$\bar\mathrm{s}$' we make use of the
scattering states
\begin{eqnarray}
   &&\hat\Psi_\mathrm{s}\!=\!\sum\limits_{k\sigma}
   e^{ikx}\hat
   a_{k\sigma}
   +e^{-ikx}(
   \cos\vartheta e^{-i\varphi}\hat c_{k\sigma}
   +\sin\vartheta e^{i\psi}\hat d_{k\sigma}),
   \nonumber\\
   &&\hat\Psi_{\bar \mathrm{s}}\!=\!\sum\limits_{k\sigma}
   e^{ikx}\hat b_{k\sigma}+e^{-ikx}(
   \cos\vartheta e^{i\varphi}\hat d_{k\sigma}
   -\sin\vartheta e^{-i\psi}\hat c_{k\sigma}),
   \nonumber
\end{eqnarray}
where $\hat a_{k\sigma}$, $\hat b_{k\sigma}$ denote the
annihilation operators for electrons in the source reservoirs
$\mathrm{s}$ and $\bar\mathrm{s}$ with momentum $k$ and spin
$\sigma\in\uparrow,\downarrow$ polarized along the $z$-axis 
and time evolution $\propto \exp(-i\epsilon_k t/\hbar)$,
$\epsilon_k = \hbar^2 k^2/2m$; the operators $\hat c_{k\sigma}$
and $\hat d_{k\sigma}$ annihilate electrons in the reservoirs
attached to the outgoing leads `u' and `d', respectively. Also,
we make use here of the standard parametrization of a
reflectionless four-beam splitter,
\begin{equation}
   \left(\begin{array}{c}\mathrm{u}\\
   \mathrm{d}\end{array}\right)=
   \left(
   \begin{array}{cc}
   e^{i\varphi}\cos\vartheta& -e^{i\psi}\sin\vartheta\\
   e^{-i\psi}\sin\vartheta& e^{-i\varphi}\cos\vartheta
   \end{array}
   \right)
   \left(\begin{array}{c}\mathrm{s}\\
   \bar\mathrm{s}\end{array}\right),
\end{equation}
with the angles $\vartheta\in(0,\pi/2)$, $\varphi,\psi\in(0,2\pi)$;
without loss of generality we will assume $\varphi=\psi=0$ in
what follows. The orbital part of the one-particle
correlator $G^\mathrm{xy}(x-y) \equiv G(x-y)$ takes the form
\begin{eqnarray}
   &&G_\mathrm{eq}(x)=
   \frac{\sin k_{\rm\scriptscriptstyle F}}{\pi x},
   \label{orb_eq}
   \\
   &&G_\mathrm{ex}(x)=e^{-i(k_{\rm\scriptscriptstyle F}+
   k_V)x}\,\frac{\sin k_V x}{\pi x},
   \label{orb_ex}
\end{eqnarray}
with  $k_V=k_{\rm\scriptscriptstyle F}(eV/\epsilon_{\rm
\scriptscriptstyle F})$ and $\epsilon_{\rm
\scriptscriptstyle F}$ ($k_{\rm \scriptscriptstyle F}$) 
the Fermi energy (wave vector) in the unbiased system.
The spin factors for the equilibrium and excess parts
read,
\begin{eqnarray}
   &&\chi_\mathrm{eq}^\mathrm{xx}(\sigma,\bar\sigma)=
     \langle\sigma| \bar\sigma\rangle,
     \label{spin_eq_ex}
     \\
   &&\chi_\mathrm{ex}^\mathrm{ss}(\sigma,\bar\sigma)\!=\!\langle\sigma|\!
     \uparrow\rangle \langle\uparrow\!|\bar\sigma\rangle,
     \quad \chi_\mathrm{ex}^\mathrm{\bar s\, \bar s }
     (\sigma,\bar\sigma)\!=\!\langle\sigma| \!\downarrow\rangle
     \langle\downarrow\!|\bar\sigma\rangle,
     \nonumber
\end{eqnarray}
the latter describing the injection of polarized electrons into
the leads `s' and `$\bar\mathrm{s}$'. Finally, the cross correlation
function between the source leads vanishes,
$G_{\sigma\bar\sigma}^\mathrm{\,s\bar s}(x-y)=0$, and the final
result for the excess part of the pair correlation function
between source leads reads
\begin{equation}
   \bigl[g^\mathrm{\,s\bar s}_{\vec \sigma}(x,y)\bigr]_\mathrm{ex}=
   |G_\mathrm{ex}(0)|^2\,
   \langle\sigma_1|\!\uparrow\rangle\langle\uparrow\!|\sigma_4
   \rangle\langle\sigma_2|\!\downarrow\rangle\langle\downarrow\!|
   \sigma_3\rangle.
\end{equation}
This result then describes the injection of two uncorrelated
streams of polarized electrons into the leads `s' and
`$\bar\mathrm{s}$'. Furthermore, statistical analysis
\cite{lesoviklevitov} tells that the Fermi statistics 
enforces injection into each lead of a regular stream 
of particles separated by the single-particle correlation 
time $\tau_V=\hbar/eV$. The full many body description 
then is conveniently reduced to a two-particle problem 
where the two reservoirs inject a sequence of electron
pairs residing in the wave function
$\Psi_\mathrm{in}^{\scriptscriptstyle
12}=\bigl[\phi_{\mathrm{s}\uparrow}^{\scriptscriptstyle 1}
\phi_{\bar\mathrm{s}\downarrow}^{\scriptscriptstyle 2}
-\phi_{\bar\mathrm{s}\downarrow}^{\scriptscriptstyle 1}
\phi_{\mathrm{s}\uparrow}^{\scriptscriptstyle 2} \bigr]/
\sqrt{2}$ with $\phi_{\mathrm{s}\uparrow}$
($\phi_{\bar\mathrm{s}\downarrow}$) the single-particle wave
functions associated with electrons in the upper (lower) source
lead. This wave function is a simple Slater determinant and hence
non-entangled according to~\cite{schliemann_01}.

Next, we extend the above analysis to the outgoing leads
`u' and `d'. The scattering states in the outgoing leads 
take the form
\begin{eqnarray}
   &&\hat\Psi_\mathrm{u}=\!\sum\limits_{k\sigma}
   e^{-ikx}\hat c_{k\sigma}\!+e^{ikx}
   (\cos\vartheta\, \hat a_{k\sigma}
   -\sin\vartheta\, \hat b_{k\sigma}),
   \nonumber\\
   &&\hat\Psi_\mathrm{d}=\!\sum\limits_{k\sigma}
   e^{-ikx}\hat d_{k\sigma}\!+e^{ikx}
   (\cos\vartheta\, \hat b_{k\sigma}
   +\sin\vartheta\, \hat a_{k\sigma}).
   \nonumber
\end{eqnarray}
The excess particles injected by the source leads now are mixed
in the beam-splitter and thus non-vanishing cross correlations
are expected to show up in the leads `u' and `d'. 
The one-particle correlation function assumes the
form (\ref{one_p}) with the orbital correlators (\ref{orb_eq})
and (\ref{orb_ex}) and spin correlators
\begin{eqnarray}
   &&\chi_\mathrm{eq}^\mathrm{xx}(\sigma,\bar\sigma)=\langle\sigma|
   \bar\sigma\rangle,\qquad \mathrm{x}\in
   {\text{`u'}},{\text{`d'}},
   \\
   &&\chi_\mathrm{ex}^\mathrm{uu}(\sigma,\bar\sigma)=
   \cos^2\vartheta\langle\sigma|\!\uparrow\rangle
   \langle\uparrow\!|\bar\sigma\rangle+
   \sin^2\vartheta\langle\sigma|\!\downarrow\rangle
   \langle\downarrow\!|\bar\sigma\rangle,
   \nonumber\\
   &&\chi_\mathrm{ex}^\mathrm{dd}(\sigma,\bar\sigma)=
   \sin^2\vartheta\langle\sigma|\!\uparrow\rangle
   \langle\uparrow\!|\bar\sigma\rangle+
   \cos^2\vartheta\langle\sigma|\!\downarrow\rangle
   \langle\downarrow\!|\bar\sigma\rangle,
   \nonumber\\
   &&\chi_\mathrm{ex}^\mathrm{ud}(\sigma,\bar\sigma)=
   \chi_\mathrm{ex}^\mathrm{du}(\sigma,\bar\sigma)
   \nonumber\\
   &&\qquad\qquad\,\, =
   \cos\vartheta\sin\vartheta
   \bigl[\langle\sigma|\!\uparrow\rangle
   \langle\uparrow\!|\bar\sigma\rangle-
   \langle\sigma|\!\downarrow\rangle
   \langle\downarrow\!|\bar\sigma\rangle\bigr].
   \nonumber
\end{eqnarray}
Evaluating the excess part of the two-particle cross-correlations
between the leads `u' and `d' at the symmetric position $x=y$
we find
\begin{eqnarray}
   &&\bigl[g^\mathrm{ud}_{\vec \sigma}(x,x)\bigr]_\mathrm{ex}
   =|G_\mathrm{ex}(0)|^2\label{dm_c}\\
   &&\qquad \,\,\, \bigl[\cos^4\vartheta
   \langle\sigma_1|\!\uparrow\rangle
   \langle\uparrow\!|\sigma_4\rangle
   \langle\sigma_2|\!\downarrow\rangle
   \langle\downarrow\!|\sigma_3\rangle
   \nonumber\\
   &&\qquad+\sin^4\vartheta\langle\sigma_1|\!\downarrow\rangle
   \langle\downarrow\!|\sigma_4\rangle
   \langle\sigma_2|\!\uparrow\rangle
   \langle\uparrow\!|\sigma_3\rangle
   \nonumber\\
   &&\qquad+\cos^2\vartheta\sin^2\vartheta
   \langle\sigma_1|\!\uparrow\rangle
   \langle\uparrow\!|\sigma_3\rangle
   \langle\sigma_2|\!\downarrow\rangle
   \langle\downarrow\!|\sigma_4\rangle
   \nonumber\\
   &&\qquad+\cos^2\vartheta\sin^2\vartheta
   \langle\sigma_1|\!\downarrow\rangle
   \langle\downarrow\!|\sigma_3\rangle
   \langle\sigma_2|\!\uparrow\rangle
   \langle\uparrow\!|\sigma_4\rangle
   \bigr].
   \nonumber
\end{eqnarray}
Hence, a symmetric splitter ($\vartheta=\pi/4$) produces the
spin correlations of a triplet state $[|\chi_\mathrm{tr}^\mathrm{ud}
\rangle=|\!\uparrow\rangle_\mathrm{u} |\!\downarrow
\rangle_\mathrm{d}+$ $|\!\downarrow\rangle_\mathrm{u}
|\!\uparrow\rangle_\mathrm{d}]/\sqrt{2}$ involving two electrons
separated in different leads `u' and `d' but at equivalent
locations $x=y$. The general case with arbitrary mixing angle $\vartheta$
results in a density matrix describing a pure state involving the
superposition $|\chi_\mathrm{tr}^\mathrm{ud}\rangle+\cos2\vartheta
|\chi_\mathrm{sg}^\mathrm{ud}\rangle$ of the above triplet state
and the singlet state $[|\chi_\mathrm{sg}^\mathrm{ud} \rangle
=|\!\uparrow\rangle_\mathrm{u} |\!\downarrow\rangle_\mathrm{d}
-|\!\downarrow\rangle_\mathrm{u} |\!\uparrow\rangle_\mathrm{d}]/\sqrt{2}$.
The analoguous calculation for the two-particle density matrix
describing electrons in the same outgoing lead `x' equal 'u' or
'd' points to the presence of singlet correlations,
\begin{eqnarray}
   &&\bigl[g^\mathrm{xx}_{\vec \sigma}(x,y)\bigr]_\mathrm{ex}
   =|G_\mathrm{ex}(0)|^2
   \langle\sigma_1|\sigma_4\rangle\langle\sigma_2|\sigma_3
   \rangle
   \label{dm_xx}
   \\
   &&\qquad\qquad-\,|G_\mathrm{ex}(x-y)|^2
   \langle\sigma_1|\sigma_3\rangle\langle\sigma_2|\sigma_4\rangle.
   \nonumber
\end{eqnarray}

Again, the above results can be used to reduce the problem from
its many-body form to a two-particle problem. Given the incoming
Slater determinant $\Psi_\mathrm{in}^{\scriptscriptstyle 12}$ we
obtain the scattered state $\Psi_\mathrm{out}^{\scriptscriptstyle
12}$ through the transformation $\phi_{\mathrm{s}\uparrow}
\rightarrow \cos\vartheta\, \phi_{\mathrm{u}\uparrow}+
\sin\vartheta\, \phi_{\mathrm{d}\uparrow}$ describing scattered
spin-$\uparrow$ electrons originating from the source lead `s' and
$\phi_{\bar\mathrm{s}\downarrow}\rightarrow -\sin\vartheta\,
\phi_{\mathrm{u}\downarrow}+ \cos\vartheta\,
\phi_{\mathrm{d}\downarrow}$ for excess spin-$\downarrow$
electrons from `$\bar\mathrm{s}$' (the wave functions
$\phi_{\mathrm{x}\sigma}= \phi_\mathrm{x}\chi_\sigma$ describe
electrons with orbital (spin) wave function $\phi_\mathrm{x}$
($\chi_\sigma$) propagating in lead `x'). The resulting
scattering wave function has the form
\begin{eqnarray}
   &&\Psi_\mathrm{out}^{\scriptscriptstyle 12}=
   \sin\vartheta\cos\vartheta\bigl[
   \phi_\mathrm{u}^{\scriptscriptstyle 1}
   \phi_\mathrm{u}^{\scriptscriptstyle 2}
   \chi_\mathrm{sg}^{\scriptscriptstyle 12}
   -\phi_\mathrm{d}^{\scriptscriptstyle 1}
   \phi_\mathrm{d}^{\scriptscriptstyle 2}
   \chi_\mathrm{sg}^{\scriptscriptstyle 12}
   \bigr]
   \nonumber\\
   &&\qquad\>\>
   +\Phi_\mathrm{ud}^{\scriptscriptstyle 12}
   \chi_\mathrm{tr}^{\scriptscriptstyle 12}
   +\cos2\vartheta\,
   \bar{\Phi}_\mathrm{ud}^{\scriptscriptstyle 12}
   \chi_\mathrm{sg}^{\scriptscriptstyle 12},
   \label{scat}
\end{eqnarray}
where the first two terms describe the propagation of
a spin-singlet pair with the wave function
$\chi_\mathrm{sg}^{\scriptscriptstyle 12}=
(\chi_\uparrow^{\scriptscriptstyle 1}
\chi_\downarrow^{\scriptscriptstyle 2}
-\chi_\downarrow^{\scriptscriptstyle 1}
\chi_\uparrow^{\scriptscriptstyle 2})/\sqrt{2}$ in the upper and
lower lead. The last two terms describe the component where
the electron pair is split between the `u' and `d' leads;
it is a superposition of singlet- and triplet states
($\chi_\mathrm{tr}^{\scriptscriptstyle 12}=
(\chi_\uparrow^{\scriptscriptstyle 1}
\chi_\downarrow^{\scriptscriptstyle 2}
+\chi_\downarrow^{\scriptscriptstyle 1}
\chi_\uparrow^{\scriptscriptstyle 2})/\sqrt{2}$) with
corresponding symmetrized and anti-symmetrized orbital wave
functions $\bar{\Phi}_\mathrm{ud}^{\scriptscriptstyle
12}=(\phi_\mathrm{u}^{\scriptscriptstyle 1}
\phi_\mathrm{d}^{\scriptscriptstyle 2}
+\phi_\mathrm{d}^{\scriptscriptstyle 1}
\phi_\mathrm{u}^{\scriptscriptstyle 2})/2$  and
$\Phi_\mathrm{ud}^{\scriptscriptstyle
12}=(\phi_\mathrm{u}^{\scriptscriptstyle 1}
\phi_\mathrm{d}^{\scriptscriptstyle 2}
-\phi_\mathrm{d}^{\scriptscriptstyle 1}
\phi_\mathrm{u}^{\scriptscriptstyle 2})/2$.
The entanglement present in these wave functions 
is easily determined using the formalism developed by Schliemann
{\it et al.} \cite{schliemann_01}: The wave function associated
with a pair of electrons can be written in terms of a single-electron
basis $\{\phi_i\}$, $\Psi^{\scriptscriptstyle 12}=\sum_{ij}
\phi_i^{\scriptscriptstyle 1} w_{ij} \phi_j^{\scriptscriptstyle
2}$ where the anti-symmetric matrix $w_{ij}=-w_{ji}$ guarantees
for the proper symmetrization. The analysis simplifies drastically
for the case where the one-particle Hilbert space is four-dimensional;
then the {\it concurrence} ${\cal C}(\Psi) =8\sqrt{\det w(\Psi)}$
gives a quantitative measure for the entanglement present in the
wave function $\Psi$, ${\cal C}(\Psi) = 0$ for a non-entangled state
and  ${\cal C}(\Psi) = 1$ for a fully entangled wave function.
For our setup the one-particle basis is defined as
$\{\phi_{\mathrm{u}\uparrow}, \phi_{\mathrm{u}\downarrow},
\phi_{\mathrm{d}\uparrow}, \phi_{\mathrm{d}\downarrow}\}$
and the matrix $w(\Psi_\mathrm{out})$ describing the scattered
state (\ref{scat}) assumes the form
\begin{eqnarray}
   w^\mathrm{out}_{ij}=\!\frac1{\sqrt2}\!\!\left[\!
   \begin{array}{cccc}
   0&-{\sin2\vartheta}/2&0&\cos^2\vartheta\\
   \sin2\vartheta/2&0&\sin^2\vartheta&0\\
   0&-\sin^2\vartheta&0&\sin2\vartheta/2\\
   -\cos^2\vartheta&0&-\sin2\vartheta/2&0
   \end{array}\!
   \right]\!.
   \nonumber
\end{eqnarray}
The concurrence of the scattering state (\ref{scat}) vanishes,
hence $\Psi^\mathrm{out}$ is non-entangled and takes the form
of an elementary Slater determinant. Next, let us analyze
the concurrence of that part of the scattering wave function
to which our coincidence measurement in leads `u' and `d'
is sensitive. The component describing the two particles split
between the leads reads
$\Psi_\mathrm{ud}^{\scriptscriptstyle 12} =
\Phi_\mathrm{ud}^{\scriptscriptstyle 12}
\chi_\mathrm{tr}^{\scriptscriptstyle 12}+\cos2\vartheta\,
\bar{\Phi}_\mathrm{ud}^{\scriptscriptstyle 12}
\chi_\mathrm{sg}^{\scriptscriptstyle 12}$, cf.\ (\ref{scat}).
This projected state is described by the matrix
\begin{eqnarray}
      w^\mathrm{ud}_{ij}=\frac1{\sqrt{2}}\left[\!
      \begin{array}{cccc}
      0&0&0&\cos^2\vartheta\\
      0&0&\sin^2\vartheta&0\\
      0&-\sin^2\vartheta&0&0\\
      -\cos^2\vartheta&0&0&0
      \end{array}\!
      \right],
      \nonumber
\end{eqnarray}
from which one easily derives the concurrence ${\cal
C}(\Psi_\mathrm{ud}^{\scriptscriptstyle
12})=\sin^22\vartheta$; we conclude that the
component $\Psi_\mathrm{ud}^{\scriptscriptstyle 12}$
detected in a coincidence measurement {\it is entangled}.
Furthermore, the concurrence is equal to unity for the
symmetric splitter $\phi=\pi/4$ where we deal with a
maximally entangled triplet state (note the loss of
information about which electron (from `s' or `$\bar
\mathrm{s}$') enters the lead `u' or `d'). We conclude that
a Bell inequality test sensitive to the split part
of the wave function will exhibit violation. We 
attribute this violation to the combined action of
{\it i)} the splitter where the information on the 
identity of the particles is destroyed and the 
entangled component $\Psi_\mathrm{ud}^{\scriptscriptstyle
12}$ is `prepared' and {\it ii)} the wave function 
projection inherent in the coincidence measurement and
`realizing' the entanglement.

The Bell type setup \cite{BItest} in Fig.\ \ref{fig:setup}
measures the correlations in the spin-entangled scattered wave
function $\Psi_\mathrm{out}^{\scriptscriptstyle 12}$. It involves
the finite-time current cross-correlators $C_{{\bf a},{\bf b}}
(x,y;\tau) \equiv \langle\langle \hat I_{\bf a}(x,\tau)
\hat I_{\bf b}(y,0) \rangle\rangle$ between the spin-currents 
$\hat I_{\bf a}(x,\tau)$ projected onto directions ${\bf a}$ 
(in lead `u') and partners $\hat I_{\bf b}(y,0)$
(in lead `d') projected onto ${\bf b}$. These correlators 
enter the Bell inequality ($\bar{\bf a}$ and
$\bar{\bf b}$ denote a second set of directions)
\begin{equation}
   |E({\bf a},{\bf b})-E({\bf a},\bar{\bf b})
  + E(\bar{\bf a},{\bf b})+E(\bar{\bf a},\bar{\bf b})| \leq 2
   \label{BI}
\end{equation}
via the current difference correlators
\begin{equation}
   E({\bf a},{\bf b}) =
   \frac{\langle[\hat{I}_{\bf a}(\tau)
   -\hat{I}_{\bf -a}(\tau)]
   [\hat{I}_{\bf b}(0)
   -\hat{I}_{\bf -b}(0)]\rangle}
   {\langle[\hat{I}_{\bf a}(\tau)
   +\hat{I}_{\bf -a}(\tau)]
   [\hat{I}_{\bf b}(0)
   +\hat{I}_{\bf -b}(0)]\rangle}.
   \label{E}
\end{equation}
The cross-measurement in different leads implies that 
the setup is sensitive only to the spin-entangled 
split-pair part $\Psi_\mathrm{ud}^{\scriptscriptstyle
12}$ of the scattering wave function and hence the Bell 
inequality can be violated. 
Making use of the field operators $\hat\Psi_\mathrm{u}$ and
$\hat\Psi_\mathrm{d}$ describing the scattering states
in the outgoing leads, we determine the irreducible current 
cross correlator and factorize into orbital and spin parts,
$C_{{\bf a},{\bf b}}(x,y;\tau) =C_{x,y}(\tau)F_{{\bf a},{\bf b}}$,
with $F_{{\bf a},{\bf b}}$ accounting for the spin projections. 
Using standard scattering theory of noise \cite{noise}, one 
obtains the orbital cross-correlator (only the excess part 
gives a finite contribution)
\begin{equation}
   C_{x,y}(\tau)
   =-\frac{e^2\sin^2\!2\vartheta}{h^2}
   \sin^2\frac{eV(\tau\!-\!\tau_-)}{\hbar}
   \,\alpha(\tau\!-\!\tau_-,\theta),
   \label{Ceqex}
\end{equation}
with $\alpha(\tau, \theta) = \pi^2\theta^2/\sinh^2 [\pi\theta\tau
/\hbar]$, $\tau^{\pm}= (x\pm y)/v_{\rm\scriptscriptstyle F}$, 
$\theta$ the temperature of the electronic reservoirs, and
$v_{\rm\scriptscriptstyle F}$ the Fermi velocity. In order to
arrive at the result (\ref{Ceqex}) we have dropped terms
small in the parameter $|\epsilon^\prime-\epsilon|/\epsilon_{\rm 
\scriptscriptstyle F}$ \cite{noise}. The spin projection 
$F_{{\bf a},{\bf b}}$ assumes the form
\begin{eqnarray}
   &&F_{{\bf a},{\bf b}}=\langle{\bf a}|\!\uparrow\rangle
   \langle\uparrow\!|{\bf b}\rangle\langle{\bf b}|\!
   \uparrow\rangle\langle\uparrow\!|{\bf a}\rangle
   +\langle{\bf a}|\!\downarrow\rangle
   \langle\downarrow\!|{\bf b}\rangle\langle{\bf b}|\!
   \downarrow\rangle\langle\downarrow\!|{\bf a}\rangle
   \nonumber\\
   &&\qquad\>-\,\langle{\bf a}|\!\uparrow\rangle
   \langle\uparrow\!|{\bf b}\rangle\langle{\bf b}|\!
   \downarrow\rangle\langle\downarrow\!|{\bf a}\rangle
   -\langle{\bf a}|\!\downarrow\rangle
   \langle\downarrow\!|{\bf b}\rangle\langle{\bf b}|\!
   \uparrow\rangle\langle\uparrow\!|{\bf a}\rangle.
   \nonumber
\end{eqnarray}
We express this result in terms of the angles $\theta_{\bf a}$ and 
$\varphi_{\bf a}$ describing the direction of magnetization in the 
`u' lead filters and $\theta_{\bf b}$, $\varphi_{\bf b}$ referring
to the filters in the `d' lead and find that $F_{{\bf a},{\bf b}}=
F_{-{\bf a},-{\bf b}}=F^+_{{\bf a},{\bf b}}$, $F_{-{\bf a},{\bf b}}
=F_{{\bf a},-{\bf b}}=F^-_{{\bf a},{\bf b}}$ and
\[
   F_{{\bf a},{\bf b}}^\pm\!=\!(1\!\pm \cos\theta_{\bf a}\cos\theta_{\bf b}
   \!\mp\!\cos\varphi_{{\bf a}{\bf b}}
   \sin\theta_{\bf a}\sin\theta_{\bf b})/2,
\]
with $\varphi_{{\bf a}{\bf b}}=\varphi_{\bf a}-\varphi_{\bf b}$.
The correlator $E({\bf a},{\bf b})$ takes the form
\begin{eqnarray}
   E({\bf a},{\bf b})=
   \frac{2C_{x,y}(\tau)\bigl[F_{{\bf a},{\bf b}}^+
   -F_{{\bf a},{\bf b}}^-\bigr]+\Lambda_-}{
   2C_{x,y}(\tau)\bigl[F_{{\bf a},{\bf b}}^++
   F_{{\bf a},{\bf b}}^-\bigr]+\Lambda_+},
   \nonumber
\end{eqnarray}
with $\Lambda_\pm = [\langle\hat{I}_{{\bf a}}\rangle \pm
\langle\hat{I}_{{\bf -a}}\rangle] [\langle\hat{I}_{{\bf b}}
\rangle\pm\langle \hat{I}_{{\bf -b}}\rangle]$. Evaluating
the projected current averages one obtains
$\Lambda_-= -e^2\, (2eV/h)^2 \cos\theta_{\bf a}
\cos\theta_{\bf b}\cos^22\vartheta$ and $\Lambda_+=e^2(2eV/h)^2$.
The triplet state is rotationally invariant within the plane
$\theta_{\bf a}=\theta_{\bf b}=\pi/2$ and choosing filters
within this equatorial plane the Bell inequality takes the form
\[
   \Bigg|\frac{C_{x,y}(\tau)
   [\cos\varphi_{{\bf a}{\bf b}}
   \!-\cos\varphi_{{\bf a}\bar{\bf b}}
   \!+\cos\varphi_{\bar{\bf a}{\bf b}}
   \!+\cos\varphi_{\bar{\bf a}\bar{\bf b}}\bigr]}
   {2 C_{x,y}(\tau)+\Lambda_+}\Bigg| \!\leq  1.
\]
Its maximum violation is obtained for the set of angles
$\varphi_{\bf a}=0$, $\varphi_{\bf b}=\pi/4$, 
$\varphi_{\bar{\bf a}}=\pi/2$, $\varphi_{\bar{\bf b}}=3\pi/4$,
\begin{equation}
   E_{\rm\scriptscriptstyle BI} \equiv
   \left|\frac{2C_{x,y}(\tau)}{
   2C_{x,y}(\tau)
   +\Lambda_+} \right|
   \leq\frac1{\sqrt{2}}.
   \label{BI_max}
\end{equation}
Evaluating the above expression in the limit of low 
temperatures $\theta < eV$ and at the symmetric position 
$x=y$, we arrive at the simple form 
\begin{equation}
   \frac{\sin^22\vartheta\,\sin^2
   (eV\tau/\hbar\bigr)}{2(eV\tau/\hbar)^2
   -\sin^22\vartheta\sin^2(eV\tau/\hbar)} \leq \frac1{\sqrt{2}}.
\label{Bell_simple}
\end{equation}
We observe that the violation of the Bell inequality is restricted
to short times $\tau < \tau_{\rm\scriptscriptstyle BI} =
\tau_V\equiv \hbar/eV$ (\cite{lebedev_03}; the relevance 
of a coincidence measurement involving the short time 
$\tau_V$ was noticed in Refs.\ \cite{beenakker_03,samuelsson_03}). 
Furthermore, the violation strongly depends on the mixing angle 
$\vartheta$ of the beam splitter, with a maximal violation
realized for a symmetric splitter $\vartheta=\pi/4$
generating a pure triplet state across the two arms. The
Bell inequality cannot be violated for asymmetric splitters
with $|\vartheta - \pi/4| > 0.2135$ (corresponding to an angular
width $|\vartheta - 45^\circ| > 12.235^\circ$): evaluating the BI 
(\ref{Bell_simple}) at zero time difference (i.e., in a 
coincidence measurement) we find the condition
\begin{equation}
  \frac{\sin^2 2\vartheta}{2-\sin^2 2\vartheta}
   \leq\frac1{\sqrt{2}},
  \label{vartheta_c}
\end{equation}
from which one derives the critical angle $\vartheta_c = [{\rm
arcsin}(2/(\sqrt{2}+1))^{1/2}]/2 = 0.572$ (or $\vartheta_c =
32.765^\circ$). The appearance of a critical angle naturally 
follows from the fact that the measured wave function component
$\Psi_\mathrm{ud}^{\scriptscriptstyle 12}$ assumes the form
of a simple Slater determinant in the limits $\vartheta=0,\pi/2$
and hence is not entangled. Note that the product 
of average currents $\Lambda_+$ is the largest
term in the denominator of (\ref{BI_max}) and hence always
relevant. 

In conclusion, we have described a mesoscopic setup with
a source injecting non-entangled electron pairs into two
source leads `s' and `$\bar\mathrm{s}$'. Subsequent 
mixing of these particle streams in a four-channel beam 
splitter does not generate entanglement between the 
particles in the two output leads `u' and `d'. However,
proper mixing of the incoming beams in the splitter removes
the information on the path of the incoming particles
and generates a  wave function component describing 
electrons split between the leads `u' and `d' which 
is entangled. It is this component which manifests
itself in the coincidence measurement of a Bell inequality 
test and proper violation is observed at short times. This
analysis answers the question regarding the origin of 
entanglement observed in the Bell inequality test applied
to the present non-interacting system. A modified
setup where the particles propagate downstream 
after a coincidence measurement lends itself as a
source for spin-entangled particles, cf.\ Ref.\
\onlinecite{bosehome_02}.

It is interesting to analyze the setup described in Ref.\
\onlinecite{lebedev_03} in the light of the findings
reported here. The setup in \cite{lebedev_03} involves
a simple normal reservoir injecting pairs of electrons
into a source lead which are subsequently separated in space
by a beam splitter. The injected pairs reside in a spin-singlet
state involving the identical orbital wave function,
$\Psi_\mathrm{in}^{\scriptscriptstyle 12} 
= \phi_\mathrm{s}^{\scriptscriptstyle 1} 
\phi_\mathrm{s}^{\scriptscriptstyle 2} 
\chi_\mathrm{sg}^{\scriptscriptstyle 12}$; the
entanglement observed in a Bell inequality test then has been
attributed to the entanglement associated with this spin-singlet
state. One may criticise, that this incoming singlet,
being a simple Slater determinant, is not entangled according 
to the definition given by Schliemann {\it et al.} 
\cite{schliemann_01}. However, after the beam splitter 
the orbital wave function $\phi_\mathrm{s}$ is 
delocalized between the two leads, 
$\phi_\mathrm{s} \rightarrow \Phi =
t_\mathrm{su}\phi_\mathrm{u}+t_\mathrm{sd}\phi_\mathrm{d}$,
with $t_\mathrm{su}$ and $t_\mathrm{sd}$ the corresponding 
scattering amplitudes. While the scattered state remains 
a Slater determinant $\Psi_\mathrm{out}^{\scriptscriptstyle 12} 
= \Phi^{\scriptscriptstyle 1} \Phi^{\scriptscriptstyle 2}
\chi_\mathrm{sg}^{\scriptscriptstyle 12}$, the singlet 
correlations now can be observed in a coincidence 
measurement testing the cross-correlations between the 
leads `u' and `d'. Hence the spin-entanglement is produced
by the reservoir, but its observation requires proper
projection. It is then difficult to trace a unique origin
for the entanglement manifested in the violation of a Bell
inequality test. The appropriate setup to address this 
question should involve a reservoir injecting
particles with opposite spin residing in a Slater determinant
of the form $\Psi_\mathrm{in}^{\scriptscriptstyle 12} 
= [\phi_{\mathrm{s}\uparrow}^{\scriptscriptstyle 1} 
\bar{\phi}_{\mathrm{s}\downarrow}^{\scriptscriptstyle 2}
- \bar{\phi}_{\mathrm{s}\downarrow}^{\scriptscriptstyle 1} 
\phi_{\mathrm{s}\uparrow}^{\scriptscriptstyle 2}]/\sqrt{2}$
which is not entangled in the spin variable.
Such an analysis has been presented here with the result,
that the orbital projection in the coincidence measurement
is sufficient to produce a spin-entangled state.

We acknowledge discussions with Atac Imamoglu and 
financial support from the Swiss National Foundation 
(SCOPES and CTS-ETHZ), the Landau Scholarship
of the FZ J\"ulich, the Russian Science Support Foundation,
the Russian Ministry of Science, and the program `Quantum
Macrophysics' of the RAS.

\end{document}